# A New Method of Strong-Coupling Expansion

Z. Y. Wang[1,*], B. He[2] and C. D. Xiong[1]

[1]University of Electronic Science and Technology of China, Chengdu, Sichuan 610054,

People's Republic of China

[2]Department of Physics, Hunter College, City University of New York, 695 Park Avenue, New York, NY 10021,USA

## Abstract

In this paper, we propose a new method of strong-coupling expansion.

As well known, for $x = 1 - ax$ ( $|a| \neq 1$ ), when $|a| < 1$, using the iterative mathematical technique we have

$$x = 1 - ax = 1 - a(1 - ax) = \cdots = 1 - a + a^2 - a^3 + \cdots = \frac{1}{1+a} \qquad (1)$$

when $|a| > 1$, we start from another form of $x = 1 - ax$, that is $x = \frac{1}{a} - \frac{x}{a}$, applying the iterative operation we have

$$x = \frac{1}{a} - \frac{1}{a}(\frac{1}{a} - \frac{x}{a}) = \cdots = \frac{1}{a} - \frac{1}{a^2} + \frac{1}{a^3} - \frac{1}{a^4} + \cdots = \frac{1}{1+a} \qquad (2)$$

On the other hand, as we known, when one calculates $I_n = e^{-1} \int_0^1 x^n e^x dx$

($n = 1,2,3,...k$), the method making use of

$$I_n = 1 - nI_{n-1} \qquad (n = 1,2,3,...k) \qquad (3)$$
$$I_0 = 1 - e^{-1}$$

is not as good as the one making use of

$$I_{n-1} = \frac{1}{n}(1 - I_n) \qquad (n = k, k-1, k-2,...,1) \qquad (4)$$
$$I_k = C_k$$

where $C_k$ is the approximate estimate of $I_k$, we call Equ. 4 as the inverse recursion of Equ. 3 with respect to $n$. Similarly, when we make use of the perturbation technique to carry out some calculation in the quantum field theory, we can choose the different forms of the related expression as the starting point of iterative expansion, where the usual expansion corresponds to the weak-coupling one. In the following we only discuss the case of strong-coupling expansion.

Let $H_i(t)$ stand for the interaction Hamiltonian, $u(t,t_0)$ the time-evolution operator from the time $t_0$ to $t$ with $u(t_0,t_0) = 1$. In general, we have [1]

$$i\frac{\partial}{\partial t}u(t,t_0) = H_i(t)u(t,t_0) \qquad (5)$$

$$u(t,t_0) = 1 - i\int_{t_0}^{t} dt_1 H_i(t_1)u(t_1,t_0) \qquad (6)$$

As well known, the weak-coupling expansion of Equ. 6 can be written as

$$u(t,t_0) = \sum_n \frac{(-i)^n}{n!} \int_{t_0}^{t} dt_1 dt_2 \cdots dt_n T[H_i(t_1)H_i(t_2)\cdots H_i(t_n)] \qquad (7)$$

where $T$ stands for the time-ordered product. In order to study the strong-coupling expansion of Equ. 6, firstly we give the following Theorem [2]:

Theorem (Weighted Mean Value Theorem for Integrals): If $f(x)$ is a function continuous on the closed interval $[a,b]$, $g(x)$ is integrable on $[a,b]$, and $g(x) \geq 0$, then there exists a number $c$, $a < c < b$, such that:

$$\int_a^b f(x)g(x)dx = f(c)\int_a^b g(x)dx \qquad (8)$$

Now we assume $H_i(t)$ is integrable and positive definite on $(-\infty,+\infty)$, $u(t,t_0)$ is continuous on $(-\infty,+\infty)$, using Equ. 8 we have

$$\int_{t_0}^{t_k} H_i(t)u(t,t_0)dt = u(t_{k+1},t_0)\int_{t_0}^{t_k} H_i(t)dt \quad (t_0 \leq t_{k+1} \leq t_k) \qquad (9)$$

Then starting from Equ. 6 we can obtain

$$u(t_n,t_0) = \sum_{j=1}^{n-1}(-1)^{j-1}\prod_{k=1}^{j}L^{-1}(t_{n-k},t_0) + (-1)^{n-1}\prod_{k=1}^{n-1}L^{-1}(t_{n-k},t_0)u(t_1,t_0)$$

$$L(t_j,t_0) = i\int_{t_0}^{t_j} H_i(t)dt \quad (j=1,2,...,n-1) \qquad (10)$$

where

$$\prod_{k=1}^{j} L^{-1}(t_{n-k},t_0) = L^{-1}(t_{n-1},t_0)L^{-1}(t_{n-2},t_0)...L^{-1}(t_{n-j},t_0),$$

$L^{-1}$ is the inverse of $L$, $u(t_1,t_0)$ can be determined by experiment or an aforehand choice, $t_0 \leq t_n \leq t_{n-1} \leq ... \leq t_2 \leq t_1$. That is to say, in contrast to the traditional procedure, here we deduct the past behavior from the present conclusion.

Obviously, Equ. 10 is to Equ. 7 as Equ. 2 is to Equ. 1, or as Equ. 4 is to Equ. 3, then we call Equ. 10 as the inverse recursion of Equ. 7 with respect to the time $t$. Let $g$ stands for the coupling constant, Equ. 7 is expanded with respect to $g$ (as $g<1$), while Equ. 10 is expanded with respect to $1/g$ (as $g>1$).

Up to now, all our demonstration is based on the first principles and do not resort to any heuristic argument.

However, Equ. 10 involves the inverse calculation and the determination of the time $t_k$ ($k=2,3,...n$). In order to see what may happen, in the practical calculation, perhaps we have to make some extreme assumption. Especially, we expect this way *happen to* be approximately valid for QCD (in our next paper we shall give a specific illustration).